\documentclass[preprint,5p,twocolumn,10pt]{elsarticle}

\usepackage[pdftex,hypertexnames=false,linktocpage=true]{hyperref}
\hypersetup{
  pdfauthor={Lin Bolang},
  colorlinks=true,
  linkcolor=blue,
  citecolor=blue,
  urlcolor=blue,
  anchorcolor=darkgreen,
  filecolor=darkgreen,
  bookmarksnumbered=true,
  pdfview=FitB
}
 \usepackage{amssymb,amsthm}
\usepackage{mathtools,cuted}
\usepackage{graphicx}
\usepackage{float}
\usepackage{subfig}
\usepackage{array}
\usepackage{booktabs}
\usepackage{arydshln}
\usepackage{multirow}
\usepackage{makecell}
\usepackage{comment}
\usepackage{braket}
\usepackage{lineno}
\usepackage{microtype}
\usepackage[section]{placeins}
\usepackage{stfloats}
\newcommand{\be}{\begin{equation}}
\newcommand{\ee}{\end{equation}}

\newcommand{\ba}{\begin{align}}
\newcommand{\ea}{\end{align}}

\begin{document}
\begin{frontmatter}

\title{Pion parton distribution functions and pion-nucleus induced $J/\psi$ production in extended light-front holographic QCD}

\author[1,2]{Jiangshan Lan\corref{cor1}}
\ead{jiangshanlan@impcas.ac.cn}

\author[1,2]{Satvir Kaur\corref{cor1}}
\ead{satvir@impcas.ac.cn}

\author[1,2]{Chandan Mondal\corref{cor1}}
\ead{mondal@impcas.ac.cn}

%\address[1]{Lanzhou University, Lanzhou 730000, China}
\address[1]{State Key Laboratory of Heavy Ion Science and Technology, Institute of Modern Physics, Chinese Academy of Sciences, Lanzhou 730000, China}
\address[2]{School of Nuclear Science and Technology, University of Chinese Academy of Sciences, Beijing 100049, China}
%\address[4]{CAS Key Laboratory of High Precision Nuclear Spectroscopy, Institute of Modern Physics, Chinese Academy of Sciences, Lanzhou 730000, China}

\cortext[cor1]{Corresponding author} 
%%abstract

\begin{abstract}

We determine the pion parton distribution functions (PDFs) from its light-front wave functions, obtained using the holographic Schr\"odinger  equation of light-front chiral QCD combined with the ’t Hooft equation in (1+1)-dimensional QCD at large $N_c$. We analyze the large-$x$ behavior of the valence PDF, $\sim (1-x)^{\beta^{\rm eff}_v}$, finding overall consistency with global analyses. These pion PDFs, together with nuclear PDFs, are then used to compute the differential cross sections up to next-to-leading order for inclusive $J/\psi$ production in pion–nucleus collisions, which show good agreement with experimental data across different energies and nuclear targets.

\end{abstract}

\end{frontmatter}

\section{Introduction}\label{intro}
The pion has a special place in quantum chromodynamics (QCD)~\cite{Callan:1977gz} as the lightest hadron and the Nambu–Goldstone boson of dynamical chiral symmetry breaking~\cite{Nambu:1961fr, Nambu:1961tp}. Its internal structure, described in terms of parton distribution functions (PDFs), provides direct insight into the underlying quark-gluon dynamics and forms an important bridge between theoretical descriptions and high-energy scattering experiments~\cite{Sutton:1991ay, Gluck:1999xe, Wijesooriya:2005ir, NA3:1983ejh, E615:1989bda, Aicher:2010cb, Barry:2018ort}.

In the framework of QCD factorization, pion PDFs encode the probability of finding a parton carrying a longitudinal momentum fraction $x$ at a given resolution scale~\cite{Peskin:1995ev}. These distributions are intrinsically scale dependent: at low scales, the pion structure is dominated by its valence degrees of freedom, while at higher scales, a richer partonic content emerges. As the resolution increases, gluon radiation and quark-antiquark pair creation dynamically generate additional partons, redistributing the pion's momentum among an increasing number of constituents.

Experimentally, pion PDFs are constrained mainly through pion-induced Drell–Yan processes, which are directly sensitive to the quark distributions, while processes such as heavy quarkonium production in pion–nucleus collisions provide complementary sensitivity to the gluon content~\cite{NA10:1987hho, Sutton:1991ay, Gluck:1999xe, Wijesooriya:2005ir}. Global analyses~\cite{Barry:2018ort,Barry:2021osv,Novikov:2020snp,Pasquini:2023aaf,Barry:2025wjx,Bourrely:2022mjf}, lattice QCD~\cite{Detmold:2003tm, Martinelli:1987bh, Oehm:2018jvm, Abdel-Rehim:2015owa, Lin:2017snn, Sufian:2019bol,Lin:2020ssv}, Dyson Schwinger and Bethe Salpeter equations (DSE/BSE)~\cite{Lan:2024ais, Yu:2024ovn, Lu:2023yna, Lu:2022cjx, Cui:2022bxn, Cui:2021mom, Chang:2021utv, Arrington:2021biu, Cui:2020dlm, Cui:2020tdf, shi:2026hqq, Aguilar:2019teb, Nguyen:2011jy}, basis light-front quantization~\cite{Lan:2021wok, Lan:2019rba}, and other phenomenological models~\cite{Bopsin:2025vhz, Ahmady:2022dfv, Kaur:2025gyr, Kaur:2020vkq, Shigetani:1993dx, Frederico:1994dx, Gutsche:2014zua, Ahmady:2018muv, deTeramond:2018ecg}, have provided valuable information on the pion structure. Nevertheless, the behavior of the valence distribution in the large-$x$ limit remains unsettled. In particular, different theoretical approaches predict distinct asymptotic behaviors, typically of the form $f_v(x, \mu^2)\sim (1-x)^{\beta_v (\mu^2)}$, where the exponent $\beta_v$ depends upon the resolution scale. While perturbative QCD predicts $\beta_v \simeq 2$~\cite{Berger:1979du, Yuan:2003fs, Farrar:1975yb}, phenomenological extractions at a fixed scale $\mu^2=1.27$ GeV$^2$ suggest a broader range, $\beta_v\sim 1.0-2.5$, depending upon the treatment of higher-order and resummation effects~\cite{Barry:2025wjx, Barry:2021osv, Barry:2018ort, Novikov:2020snp, Pasquini:2023aaf, Bourrely:2022mjf}.
%~\cite{Shigetani:1993dx, Frederico:1994dx, Melnitchouk:2002gh, Barry:2025wjx, Barry:2021osv, Barry:2018ort, Novikov:2020snp, Pasquini:2023aaf, Bourrely:2022mjf}.

From the theoretical side, a reliable description of the pion structure requires a framework that consistently incorporates confinement and relativistic bound-state dynamics. Light-front (LF) holographic QCD~\cite{Brodsky:2006uqa, deTeramond:2005su, deTeramond:2008ht} provides a promising approach in this direction, offering an effective semiclassical description of hadrons with confinement encoded in the transverse dynamics (for review article, see \cite{Brodsky:2014yha}). In recent developments, the inclusion of longitudinal dynamics through the 't~Hooft equation at large $N_c$~\cite{tHooft:1974pnl} has enabled a more complete treatment of hadronic wave functions~\cite{Ahmady:2021lsh, Ahmady:2021yzh, Ahmady:2022dfv, Gurjar:2024wpq, Gurjar:2025kcp, Kaur:2025gyr, Kaur:2025css}. Within this framework, a wide range of pion properties, including spectroscopy, decay constants, form factors, and partonic observables, have been successfully described while satisfying key constraints associated with chiral symmetry~\cite{Ahmady:2022dfv}.

Building on this established framework, in the present work we employ the pion light-front wave functions (LFWFs) to extract the pion PDFs and investigate their phenomenological implications. In particular, we focus on the large-$x$ behavior of the valence distribution and provide quantitative predictions within this approach. These PDFs are then used to study inclusive $J/\psi$ production in pion--nucleus collisions, allowing for a direct comparison with experimental data and global analyses.

\section{Light-Front Holographic Framework with Longitudinal Dynamics}\label{lfhtHooft}
%\section{Light-Front Holography and Longitudinal Dynamics}

LF holographic QCD provides an effective framework to describe hadronic structure by exploiting a correspondence between strongly coupled QCD dynamics and a higher-dimensional gravitational theory in anti-de Sitter (AdS) space~\cite{Brodsky:2006uqa, deTeramond:2005su, deTeramond:2008ht, Brodsky:2014yha}. In this approach, the bound-state problem in physical spacetime is mapped onto a semiclassical description in terms of a single-variable LF Schr\"odinger equation, allowing for an analytic treatment of confinement.

For a mesonic system, the LFWFs $\Psi(x,\mathbf{b}_\perp)$ depends on the longitudinal momentum fraction $x$ carried by the quark and the transverse separation $\mathbf{b}_\perp$ between the quark and antiquark. A key step in establishing the holographic connection is the introduction of the invariant transverse variable~\cite{Brodsky:2014yha}
\begin{equation}
\zeta = \sqrt{x(1-x)}\, b_\perp ,
\end{equation}
which provides a measure of the transverse separation and is identified with the fifth dimension of AdS space. This allows the wave function to be expressed in a factorized form separating transverse and longitudinal dynamics,
\begin{equation}
\Psi(x,\zeta,\phi) = \frac{\phi(\zeta)}{\sqrt{2\pi \zeta}} \, e^{iL\phi} \, X(x), \label{eq:lfwf}
\end{equation}
where $\phi(\zeta)$ and $X(x)$ encode the transverse and longitudinal structure, respectively, with $X(x)=\sqrt{x(1-x)}\chi(x)$.

Within this framework, the transverse dynamics is governed by an effective LF Schr\"odinger equation~\cite{Brodsky:2014yha},
\begin{equation}\label{LFH}
\left[-\frac{{\rm d}^2}{{\rm d}\zeta^2} + \frac{4L^2 - 1}{4\zeta^2} + U_\perp(\zeta)\right]\phi(\zeta) = M^2_\perp \, \phi(\zeta),
\end{equation}
where $U_\perp(\zeta)$ represents the confining potential. A quadratic form of the potential~\cite{Brodsky:2006uqa, deTeramond:2005su, deTeramond:2008ht, Brodsky:2014yha},
\begin{equation}\label{potential}
U_\perp(\zeta) = \kappa^4 \zeta^2 + 2\kappa^2 (J - 1),
\end{equation}
leads to a harmonic oscillator spectrum, with the mass scale $\kappa$ setting the strength of confinement and $J = L + S$ being the total angular momentum of the meson. This potential thus determines the hadronic spectrum.
 This formulation successfully reproduces key features of hadron spectroscopy, including linear Regge trajectories and the emergence of a massless pion in the chiral limit. With Eq.~\eqref{potential},
Eq.~\eqref{LFH} can be solved analytically, yielding the meson masses and eigenstates within the chiral limit:
\begin{equation}
	M_{\perp}^2(n_\perp , J, L)=4\kappa^2\left(n_\perp + \frac{J+L}{2}\right),
	\label{MTM}
\end{equation}
and 
\begin{equation}
	\phi_{n_{\perp} L}(\zeta )\propto \zeta^{1/2+L}\exp\left(\frac{-\kappa^2\zeta^2}{2}\right)L_{n_\perp}^L(\kappa^2\zeta^2),
	\label{TMWF}
\end{equation}
where $n_{\perp}$ is the transverse quantum number, $L_{n_\perp}^{L}$ represents the associated Laguerre polynomials.

For the ground-state pion ($n_\perp = L = 0$), the solution of the transverse LF Schr\"odinger equation leads to a Gaussian form for the wave function. In particular, the holographic LFWF can be written as
\begin{equation}
\Psi(x,\zeta) \propto \sqrt{x(1-x)} \chi(x) \,
\exp\left(-\frac{\kappa^2 \zeta^2}{2}\right),
\label{eq:lfwf-2}
\end{equation}
which exhibits confinement in the transverse direction and vanishes at the endpoints $x \to 0,1$.

While the transverse structure is dynamically determined in this approach, the longitudinal dynamics is not treated dynamically in the original formulation; instead the mode $\chi(x)=1$~\cite{Brodsky:2007hb, Brodsky:2008pf}. To achieve a more complete description of hadronic wave functions, the longitudinal mode can be constrained using the 't~Hooft equation, derived in $(1+1)$-dimensional QCD in the large-$N_c$ limit~\cite{tHooft:1974pnl, Chabysheva:2012fe, Ahmady:2021lsh}. This equation incorporates the effects of confinement in the longitudinal direction and quark masses, providing a dynamical basis for the function $\chi(x)$.

The combination of LF holography with the 't~Hooft equation thus yields a unified framework in which both transverse and longitudinal dynamics are consistently treated~\cite{Ahmady:2021lsh, Ahmady:2021yzh, Ahmady:2022dfv, Gurjar:2024wpq, Gurjar:2025kcp, Kaur:2025gyr, Kaur:2025css}. This enables the construction of realistic LFWFs that can be applied to a wide range of hadronic observables, including parton distribution functions and exclusive processes.

To incorporate dynamical information in the longitudinal direction, we employ the 't~Hooft equation, which arises in $(1+1)$-dimensional QCD in the large-$N_c$ limit. In this framework, the meson bound-state problem reduces to an effective integral equation for the longitudinal wave function $\chi(x)$~\cite{tHooft:1974pnl}, given by
\begin{equation}
\left( \frac{m_q^2}{x} + \frac{m_{\bar{q}}^2}{1-x} \right)\chi(x) + U_{\parallel}(x)\,\chi(x) = M_{\parallel}^2 \, \chi(x),
\end{equation}
where the interaction term is defined as
\begin{equation}
U_{\parallel}(x)\,\chi(x) = \frac{g^2}{\pi} \, \mathcal{P} \int_0^1 dy \, \frac{\chi(x) - \chi(y)}{(x-y)^2}.
\end{equation}
Here, $m_q$ and $m_{\bar{q}}$ denote the quark and antiquark masses, $g$ sets the strength of longitudinal confinement, and $\mathcal{P}$ indicates the Cauchy principal value prescription required to regulate the singular kernel.

In contrast to the transverse LF Schr\"odinger equation, the 't~Hooft equation does not admit closed-form analytical solutions and must be solved numerically~\cite{Chabysheva:2012fe}. We determine its eigenvalues and eigenfunctions using a matrix diagonalization method.

The resulting longitudinal wave functions exhibit a nontrivial structure governed by the interplay of quark masses and confinement. Together with Eq.~\eqref{eq:lfwf-2}, they enable the construction of the complete LFWFs using LF holography supplemented by the 't~Hooft approach. The resulting wave function satisfies the normalization condition
\begin{equation}
 \int {\rm d}x\, {\rm d}^2 \mathbf{b}_\perp \, \left| \Psi(x,\zeta) \right|^2 = 1,
\end{equation}
which is equivalently expressed in momentum space as
\begin{equation}
\int \frac{{\rm d}x\,{\rm d}^2 \mathbf{k}_\perp}{16\pi^3} \, \left| \Psi(x,\mathbf{k}_\perp) \right|^2 = 1.
\end{equation}

\section{Numerical results}
We fix the model parameters by fitting the mass spectrum of the pion family~\cite{Ahmady:2022dfv}. In particular, we use $m_{u/d} = 0.046~\text{GeV}$ for the light quark mass, $\kappa = 0.523~\text{GeV}$ for the transverse confinement scale, and $g = 0.109~\text{GeV}$ for the longitudinal confinement scale~\cite{Ahmady:2022dfv}.  

\subsection{Pion PDFs}
%\section{Pion Parton Distribution Functions}

We compute the pion PDFs starting from the LFWFs in momentum space. The valence quark distribution at the model scale is defined as
\begin{equation}
f_v^{\pi}(x,\mu_0^2) = \int \frac{d^2 \mathbf{k}_\perp}{16\pi^3} \, \left| \Psi(x,\mathbf{k}_\perp) \right|^2,
\end{equation}
which follows from the normalization of the LFWFs. At the initial scale $\mu_0^2 \approx 0.305~\text{GeV}^2$, the pion is described purely in terms of its valence degrees of freedom, with no explicit gluon or sea quark contributions.

The full set of PDFs at higher scales is obtained through QCD evolution. We evolve the distributions from the initial scale to larger momentum scales using the next-to-next-to-leading order (NNLO) Dokshitzer-Gribov-Lipatov-Altarelli-Parisi (DGLAP) evolution equations~\cite{Dokshitzer:1977sg,Gribov:1972ri,Altarelli:1977zs}, solved numerically~\cite{Salam:2008qg}. In this framework, gluons and sea quarks are generated dynamically through perturbative splitting processes. In particular, gluons arise from quark radiation processes $q \rightarrow qg$, while sea quarks are produced through gluon splitting $g \rightarrow q\bar{q}$. These coupled evolution equations redistribute momentum among partons, leading to a buildup of gluon and sea quark densities as the scale increases. Consequently, the gluon and sea distributions become significant at small $x$, while the valence distribution is gradually shifted toward lower momentum fractions.

The initial scale is fixed by matching the first moment of the valence distribution to values extracted from global analyses, namely $2\langle x \rangle_v = 0.48 \pm 0.01$ at $\mu^2 = 5~\text{GeV}^2$~\cite{Barry:2018ort}. The evolved PDFs satisfy the momentum sum rule,
\begin{equation}
\sum_{i=q,\bar{q},g} \int_0^1 dx \, x \, f_i^{\pi}(x,\mu^2) = 1,
\end{equation}
ensuring that the total momentum of the pion is shared among its constituents.

\begin{figure}[t]
\centering \includegraphics[width=.99\columnwidth]{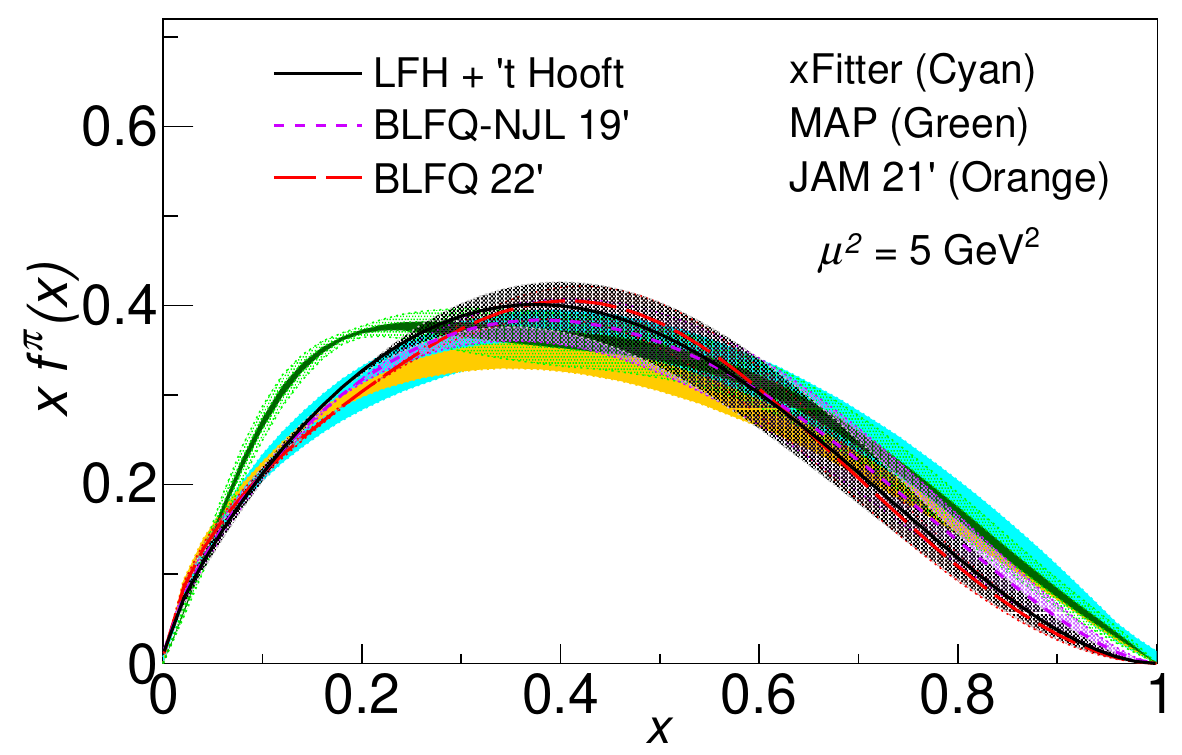}
\includegraphics[width=.99\columnwidth]{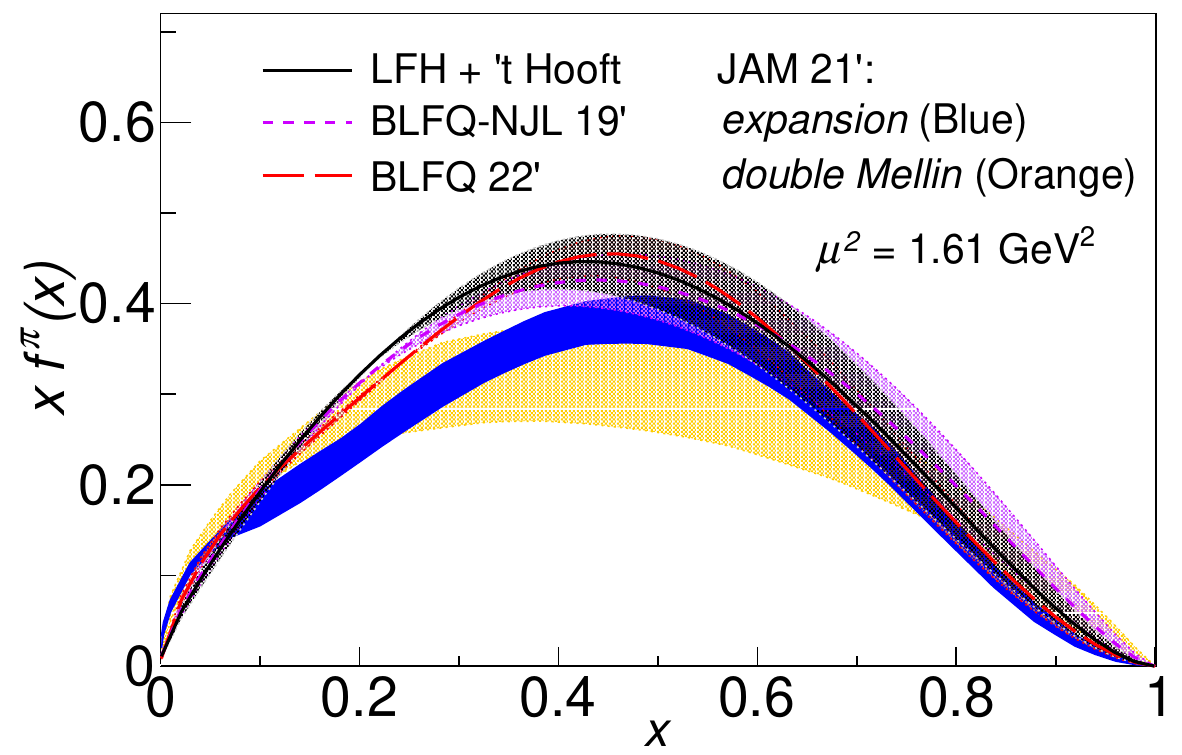}
\caption{Pion valence quark distribution $x f^{\pi}(x)$ as a function of $x$. Our results from light-front holographic QCD supplemented by the 't Hooft equation (black solid line with black band) are compared with BLFQ using an effective NJL interaction (dashed magenta line)~\cite{Lan:2019vui}, BLFQ with one dynamical gluon (long-dashed red line)~\cite{Lan:2021wok}, and global analyses by xFitter 2020 (cyan band)~\cite{Novikov:2020snp}, MAP (green band)~\cite{Pasquini:2023aaf}, and JAM 2021 (orange and blue bands)~\cite{Barry:2021osv}. The upper and lower panels show the PDFs at $5~\text{GeV}^2$ and $1.61~\text{GeV}^2$, respectively.}
\label{fpionE6152}
\end{figure}

In Fig.~\ref{fpionE6152}, we present the valence quark distribution $x f_v^\pi(x)$ at $\mu^2 = 1.61 ~\text{GeV}^2 ~{\rm and}~ 5~\text{GeV}^2$. Our results are compared with global fits from xFitter \cite{Novikov:2020snp}, MAP \cite{Pasquini:2023aaf}, and JAM \cite{Barry:2021osv}, as well as model calculations from BLFQ-NJL \cite{Lan:2019vui} and BLFQ \cite{Lan:2021wok}. The black bands represents the uncertainty in our results due to a $10\%$ uncertainty in the initial scale. The distribution exhibits the expected behavior, peaking at intermediate $x$. We find good overall agreement with these phenomenological extractions over a broad range of $x$~\cite{Novikov:2020snp, Pasquini:2023aaf, Barry:2021osv, Barry:2025wjx}.

\begin{figure}[t]
\centering \includegraphics[width=.99\columnwidth]{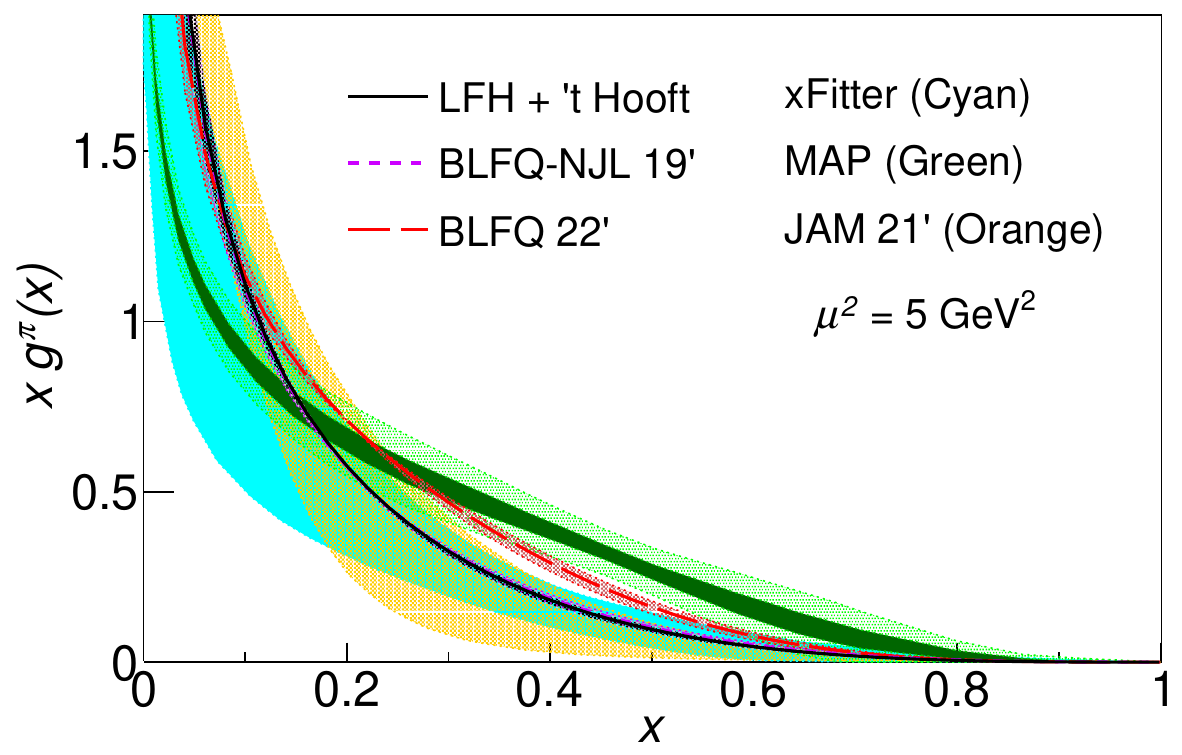}
\includegraphics[width=.99\columnwidth]{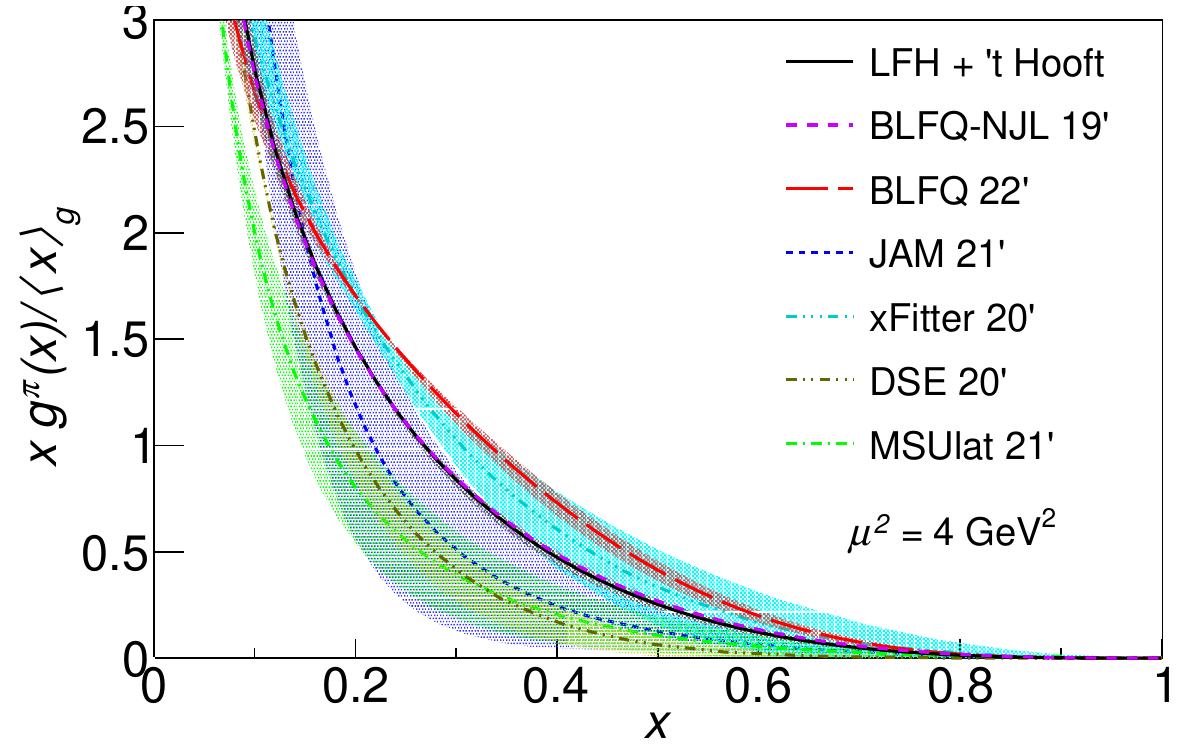}
\caption{Pion gluon distribution $x g^{\pi}(x)$  (upper) and the normalized distribution $x g^{\pi}(x)/\langle x \rangle_g$ (lower) as a function of $x$. Our results (black solid line with black band) are compared with BLFQ using an effective NJL interaction (dashed magenta line)~\cite{Lan:2019vui}, BLFQ with one dynamical gluon (long-dashed red line)~\cite{Lan:2021wok}, and global analyses by xFitter 2020 (cyan band)~\cite{Novikov:2020snp}, MAP (green band)~\cite{Pasquini:2023aaf}, and JAM 2021 (orange band and blue line)~\cite{Barry:2021osv}, as well as with DSE (brown ashed-double dot line)~\cite{Cui:2020tdf} and lattice QCD (MSULat: green dashed-dot line)~\cite{Lin:2020ssv}. The upper and lower panels show the distributions at $5~\text{GeV}^2$ and $4~\text{GeV}^2$, respectively.}
\label{fpionE6153:gluon}
\end{figure}

In Fig.~\ref{fpionE6153:gluon}, we present the pion gluon distribution $x g^\pi(x)$ (top panel) at $\mu^2 = 5~\text{GeV}^2$ and the normalized distribution $x g^\pi(x)/\langle x \rangle_g$ (bottom panel) at $\mu^2 = 4~\text{GeV}^2$. We compare our results with a range of global fits, including xFitter~\cite{Novikov:2020snp}, MAP~\cite{Pasquini:2023aaf}, and JAM~\cite{Barry:2021osv}, as well as with model predictions from BLFQ-NJL~\cite{Lan:2019vui} and BLFQ~\cite{Lan:2021wok}. For the normalized distribution, we also consider comparisons with DSE calculations~\cite{Cui:2020tdf} and lattice QCD results (MSULat)~\cite{Lin:2020ssv}.

In our framework, the gluon distribution is not introduced explicitly at the initial scale but is generated dynamically through DGLAP evolution from the valence quark distribution. As a result, it shows the expected enhancement at small $x$ and a rapid decrease at large $x$. We find good overall agreement with global analyses and other theoretical approaches over a broad range of $x$~\cite{Novikov:2020snp, Pasquini:2023aaf, Barry:2021osv, Lan:2019vui,Lan:2021wok, Cui:2020tdf, Lin:2020ssv, Kaur:2025gyr, Lan:2024ais}.

\begin{figure}[t]
\centering \includegraphics[width=.99\columnwidth]{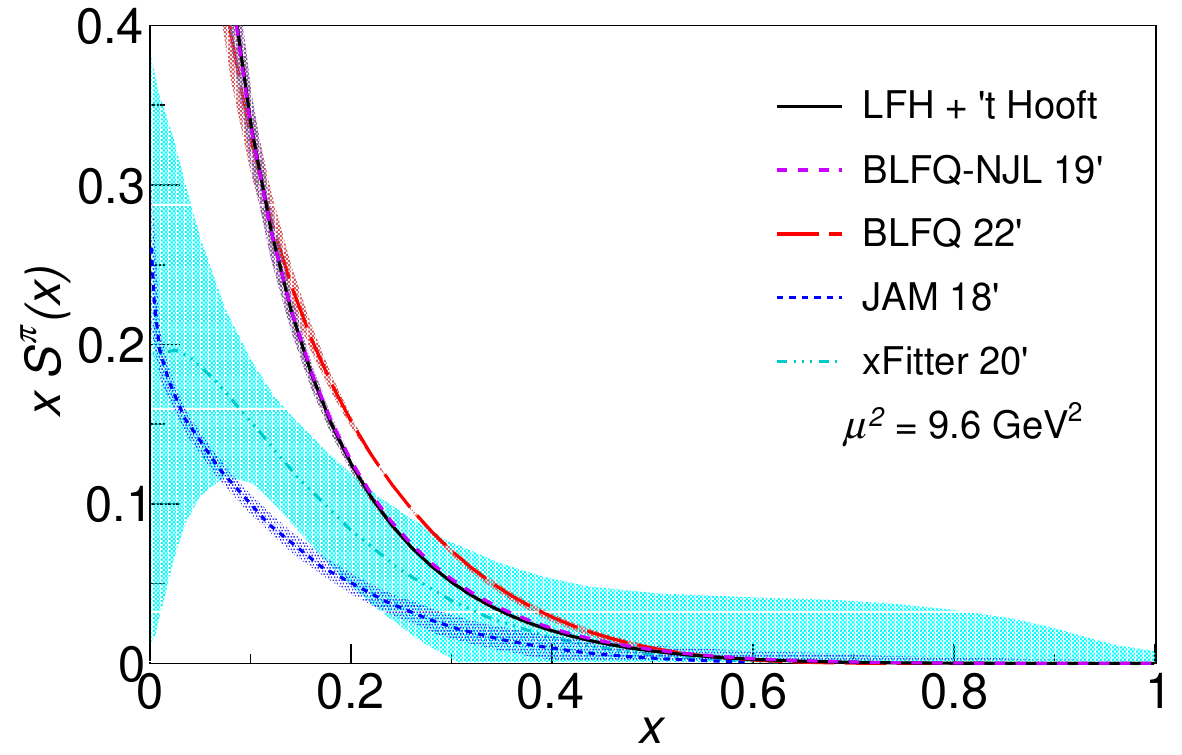}
\caption{Pion sea quark distribution $x S^{\pi}(x)$ as a function of $x$. Our results (black solid line with band) are compared with BLFQ using an effective NJL interaction (dashed magenta line)~\cite{Lan:2019vui}, BLFQ with one dynamical gluon (long-dashed red line)~\cite{Lan:2021wok}, and global analyses by JAM 2018 (blue band with dashed line)~\cite{Barry:2018ort} and xFitter 2020 (cyan band with dashed-double dot line)~\cite{Novikov:2020snp}, all at $\mu^2 = 9.6~\text{GeV}^2$.}
\label{fpionE6152:sea-quark}
\end{figure}

In Fig.~\ref{fpionE6152:sea-quark}, we show the pion sea quark distribution $x S^\pi(x)$ at $\mu^2 = 9.6~\text{GeV}^2$, compared with xFitter \cite{Novikov:2020snp} , JAM \cite{Barry:2018ort}, BLFQ-NJL \cite{Lan:2019vui}, and BLFQ \cite{Lan:2021wok}. The sea is dynamically generated through DGLAP evolution, similar to the gluon, and is concentrated at small $x$.

Overall, the extended LF holographic QCD framework, supplemented by the 't~Hooft equation, provides a consistent description of pion PDFs. The agreement with global fits and other theoretical approaches indicates that this framework captures the essential features of pion structure across a wide range of momentum fractions and energy scales.

\begin{figure*}[t]
\centering
\includegraphics[width=0.99\columnwidth]{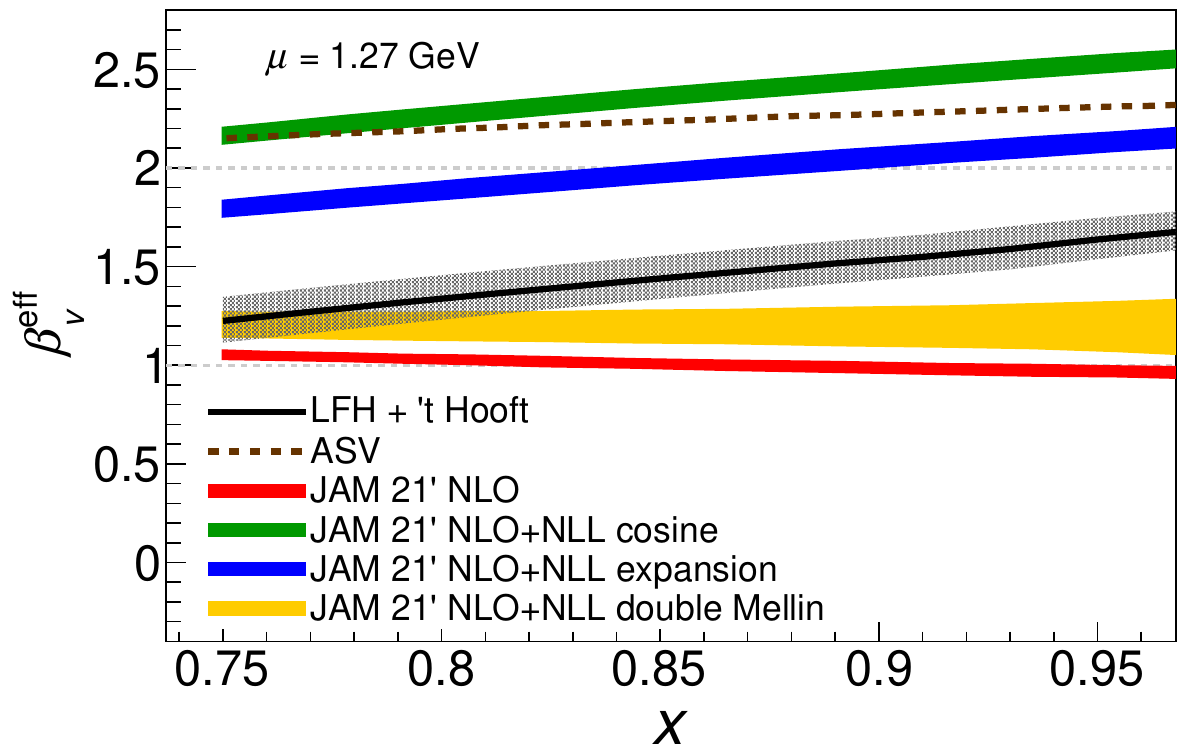}
\centering \includegraphics[width=.99\columnwidth]{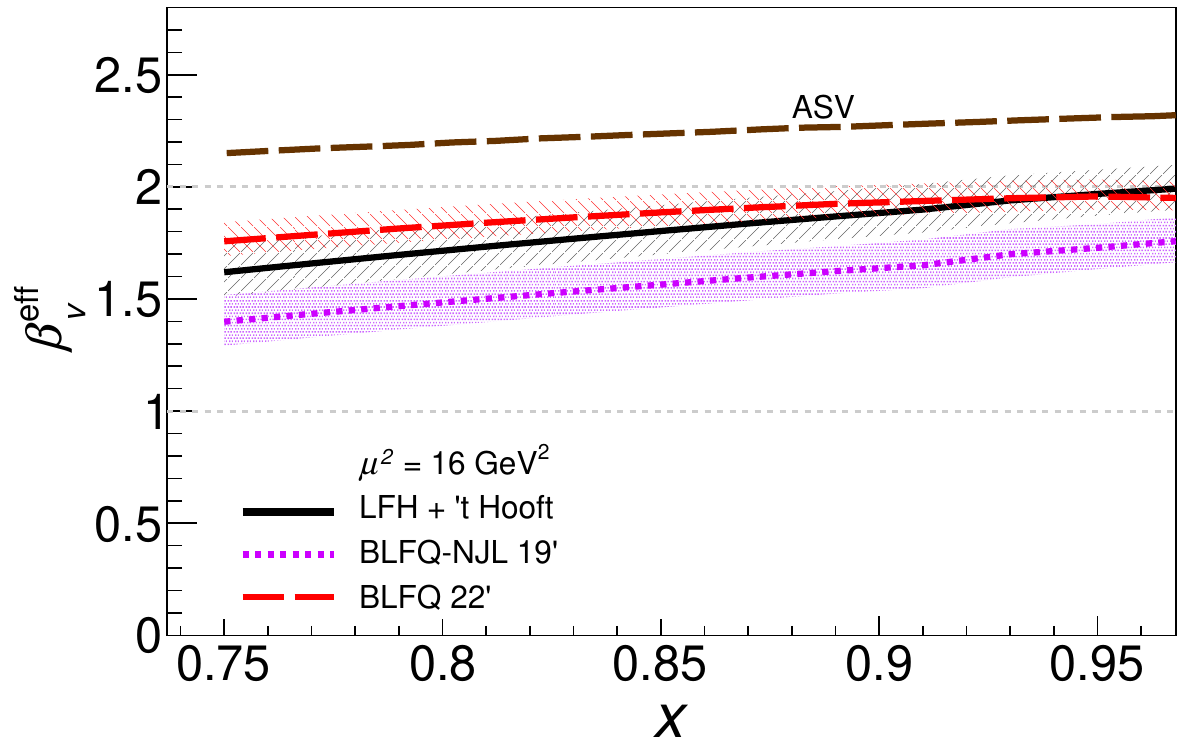}
\centering \includegraphics[width=.99\columnwidth]{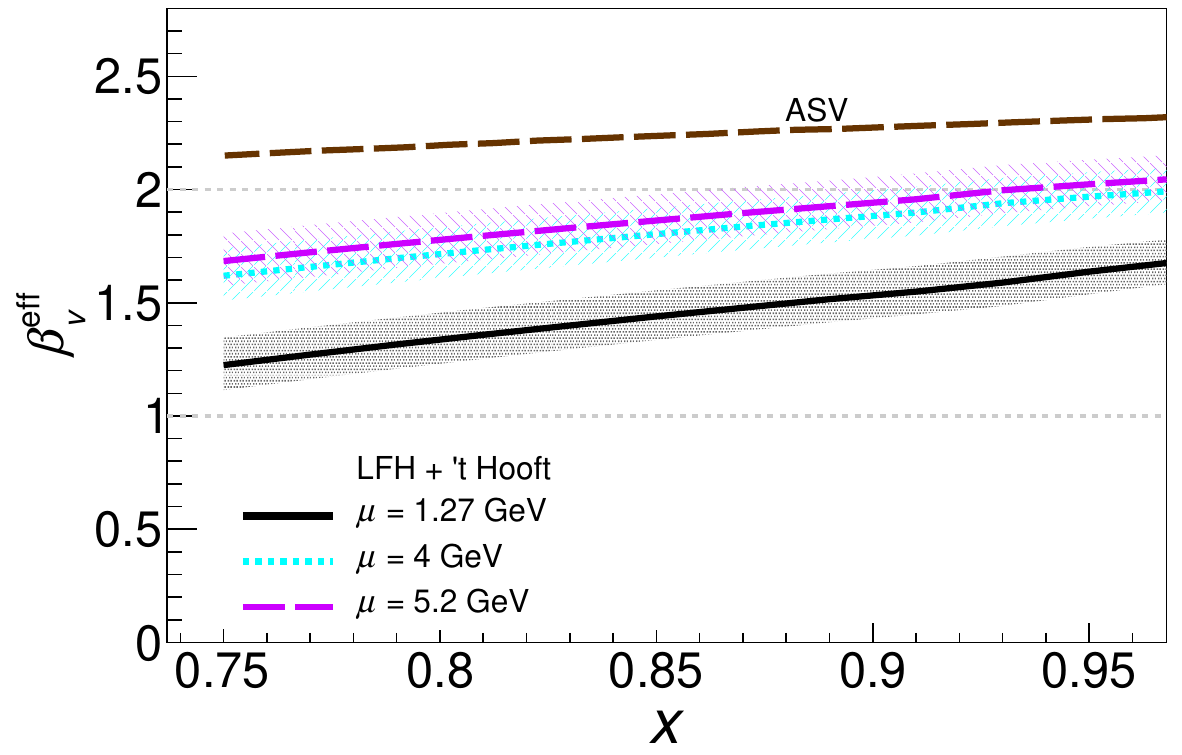}
\caption{The effective exponent $\beta_v^{\rm eff}(x)$ defined in Eq.~\eqref{beta} as a function of $x$ for the pion valence PDF at large $x$. Top left: our result (black solid line) compared with JAM21 (red, green, blue, and orange bands)~\cite{Barry:2021osv} and ASV (brown dashed line)~\cite{Aicher:2010cb} at $\mu = 1.27~\text{GeV}$. Top right: comparison with BLFQ-NJL (magenta dashed line)~\cite{Lan:2019vui} and BLFQ (red long-dashed line)~\cite{Lan:2021wok} at $\mu^2 = 16~\text{GeV}^2$. Bottom: scale dependence of our results at $\mu = 1.27~\text{GeV}$ (black line), $4~\text{GeV}$ (cyan dashed line), and $5.2~\text{GeV}$ (magenta long-dashed line).}
\label{fpionE6155-beta}
\end{figure*}

%=============================================
\subsection{Large-$x$ behavior of pion PDFs}
%=============================================
The behavior of the pion valence PDF at large $x$ can be characterized through an effective exponent defined as~\cite{Nocera:2014uea, Courtoy:2020fex, Courtoy:2021xpb, Barry:2021osv}
\begin{equation}\label{beta}
\beta_v^{\rm eff}(x,\mu^2) = \frac{\partial \ln f_v^\pi(x,\mu^2)}{\partial \ln(1-x)},
\end{equation}
which provides a local measure of the fall-off of the distribution as $x \to 1$.

In Fig.~\ref{fpionE6155-beta} (top left), we present $\beta_v^{\rm eff}$ at $\mu = 1.27~\text{GeV}$ and compare our results with JAM21~\cite{Barry:2021osv} and the ASV parametrization~\cite{Aicher:2010cb}. Our prediction (black solid line) increases steadily with $x$, reaching $\beta_v^{\rm eff} \sim 1.6\text{--}1.7$ near $x \approx 0.95$. This lies above the JAM21 next-to-leading order (NLO) result and is closer to the resummed JAM21 extractions (NLO+NLL), though still below the cosine-resummed curve. 

In Fig.~\ref{fpionE6155-beta} (top right), we compare our results with BLFQ-NJL~\cite{Lan:2019vui} and BLFQ~\cite{Lan:2021wok} at $\mu^2 = 16~\text{GeV}^2$. Our prediction remains systematically larger than BLFQ-NJL but closely follows the BLFQ results across the entire $x$ range, exhibiting a similar $x$-dependence. 

The bottom panel shows the scale dependence of $\beta_v^{\rm eff}$ for $\mu = 1.27~\text{GeV}, 4~\text{GeV},$ and $5.2~\text{GeV}$. We observe that $\beta_v^{\rm eff}$ increases with the scale, reflecting that DGLAP evolution softens the valence distribution at large $x$. The large-$x$ behavior in our framework exhibits a moderate fall-off consistent with perturbative QCD expectations and a clear scale dependence driven by QCD evolution.

\subsection{Pion-nucleus induced $J/\psi$ production}\label{pionNjpsi}
We compute the differential cross section for charmonium production up to NLO in pion--nucleus collisions and compare with existing $J/\psi$ data. We employ the Color Evaporation Model (CEM)~\cite{Einhorn:1975ua,Fritzsch:1977ay,Halzen:1977rs}, where the produced $c\bar{c}$ pair hadronizes into a specific charmonium state with a universal probability encoded in a single phenomenological parameter $F$. This approach successfully describes fixed-target and collider $J/\psi$ data~\cite{Gavai:1994in,Schuler:1996ku,Nelson:2012bc,Lansberg:2020rft}. Within this framework, the differential cross section for $J/\psi$ production in pion--nucleus interactions is given by~\cite{Nason:1987xz,Nason:1989zy,Mangano:1992kq,Chang:2020rdy,Lan:2021wok}.
\begin{align}
&\frac{d{\bf \sigma}}{dx_{\mathrm{F}}}\Big\vert_{J/\psi} =F\sum_{i,j=q,\bar{q},g}\int^{2m_D}_{2m_c}dM_{c\bar{c}}\frac{2M_{c\bar{c}}}{S\sqrt{x_F^2+4M^2_{c\bar{c}}/S}} \,\nonumber\\
&\times \hat{\bf \sigma}_{ij}(s,m^2_{c},\mu_F^2,\mu_R^2) f^{\pi^\pm}_i(x_1,\mu_F^2)\,{f}^N_{j}(x_2,\mu^2_F)\, .\label{crosseq}
\end{align}
where $x_{1,2} = (\sqrt{x_F^2 + 4M_{c\bar{c}}^2/S} \pm x_F)/2$ and $x_F = x_1 - x_2$ is the Feynman variable. Here, $S$ is the squared center-of-mass energy of the $\pi N$ system, and $s$ denotes the partonic invariant mass. The masses $m_c$, $m_D$, and $M_{c\bar{c}}$ correspond to the charm quark, $D$ meson, and the $c\bar{c}$ pair, respectively. The short-distance cross section $\hat{\sigma}_{ij}$ is calculated at NLO accuracy \cite{Nason:1987xz}. 

The hadronic inputs are the pion PDFs $f^{\pi^\pm}$ from our model and the nuclear PDFs $f^{N}$, for which we use the nCTEQ15 parametrization \cite{Kovarik:2015cma}. We choose $\mu_F = 2m_c$ and $\mu_R = m_c$, with $m_c = 1.5~\text{GeV}$.

\begin{figure}[tbp]
\centering \includegraphics[width=.99\columnwidth]{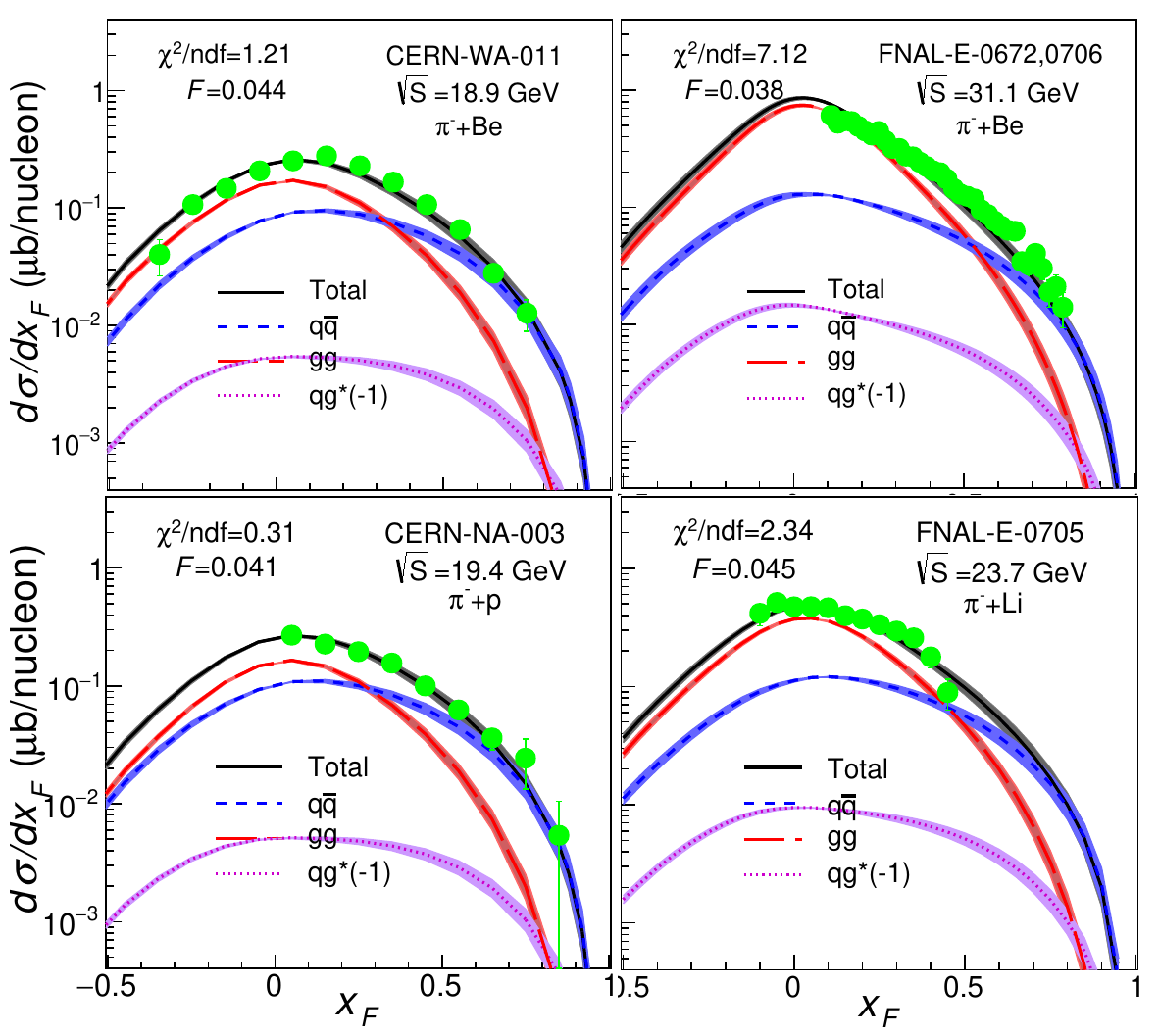}
\caption{ Differential cross section $ d\sigma/dx_{F}$ for inclusive $J/\psi$  production in $\pi^-$ nucleus collisions as a function of the Feynman variable $x_{F}$. Experimental points are compiled from the FNAL E672, E706, E705 collaborations and the CERN NA3, WA11 programs ~\cite{E672:1995won,E705:1992jno,NA3:1983ltt,McEwen:1982fe}. The $qg$  subprocess contribution, being negative-definite, is displayed here after multiplication by $-1$ to facilitate visual comparison.}
\label{fpionE6156:cross-section}
\end{figure}

In Fig.~\ref{fpionE6156:cross-section}, we compare our NLO predictions with fixed-target $J/\psi$ production data for various nuclei from E672/E706 (515 GeV $\pi^-$ + Be) \cite{E672:1995won}, E705 (300 GeV $\pi^-$ + Li) \cite{E705:1992jno}, NA3 (200 GeV $\pi^-$ + H) \cite{NA3:1983ltt}, and WA11 (190 GeV $\pi^-$ + Be) \cite{McEwen:1982fe}. The normalization factor $F$ is determined separately for each experiment via a $\chi^2$ fit, with the corresponding values and $\chi^2$/ndf shown in the figure.

At the partonic level, the $gg$ fusion channel dominates the cross section for $x_F \lesssim 0.5$, but decreases rapidly at larger $x_F$. In contrast, the $q\bar{q}$ channel provides a relatively flat contribution and becomes increasingly important at large $x_F$, reflecting the growing role of valence quarks. The $qg$ channel contributes negatively with a small magnitude. 

This behavior highlights the sensitivity of $J/\psi$ production to the pion gluon distribution, particularly in the region $x \gtrsim 0.2$. Our results show good agreement with the experimental data across different energies and nuclear targets, demonstrating that the gluon distribution obtained in our framework provides a consistent description of the underlying dynamics, in line with the findings of Ref.~\cite{Chang:2020rdy}.

\section{Conclusion}
In this work, we have investigated the pion structure within an extended light-front holographic QCD framework supplemented by the ’t~Hooft equation for longitudinal dynamics. This unified approach provides a consistent description of confinement and has been previously validated through meson spectra, electromagnetic form factors, and decay constants, thereby establishing the resulting light-front wave functions as realistic representations of pion structure. Using the resulting wave functions, we extracted the pion PDFs and evolved them using DGLAP equations, where gluon and sea quark distributions are generated dynamically. The valence, gluon, and sea PDFs are in good agreement with global analyses such as xFitter~\cite{Novikov:2020snp}, JAM~\cite{Barry:2025wjx, Barry:2021osv, Barry:2018ort}, and MAP~\cite{Pasquini:2023aaf}.

We further examined the large-$x$ behavior of the valence PDF, finding a moderate fall-off with $\beta_v^{\rm eff} \sim 1.6$--$1.7$ at $x \approx 0.95$ and a clear scale dependence. As an independent validation, we computed pion–nucleus–induced $J/\psi$ production at NLO and obtained good agreement with experimental data. The results highlight the sensitivity of this process to both gluon and valence dynamics, demonstrating that the combined light-front holographic and ’t~Hooft framework provides a consistent description of pion structure across different observables.

\section*{Acknowledgments}
This work is supported by the National Natural Science Foundation of China under Grant No. 12305095, by the Gansu International Collaboration and Talents Recruitment Base of Particle Physics (2023-2027), by the Senior Scientist Program funded by Gansu Province, Grant No. 25RCKA008,
by the Special Research Assistant Funding Project, Chinese Academy of Sciences, by China Postdoctoral Science Foundation (CPSF), Grant No. E339951SR0, and by Gansu Provincial Young Talents Program.

%%%%%%%%%%%%%%%%%%%%%%%%%%%%%%%%%%%%%%%%%%%%%%%%%%%%%%%%%%%%%%%%%%%%%
\biboptions{sort&compress}
\bibliographystyle{apsrev}
\bibliography{ref}
%%%%%%%%%%%%%%%%%%%%%%%%%%%%%%%%%%%%%%%%%%%%%%%%%%%%%%%%%%%%%%%%%%%%%

\end{document}